\documentclass[letterpaper, 10 pt, conference]{ieeeconf}  

\IEEEoverridecommandlockouts                              

\usepackage{graphics} 
\usepackage{epsfig} 
\usepackage{amsmath} 
\usepackage{amssymb}  
\usepackage{accents}
\usepackage{todonotes}

\newcommand{\ipro}[2]{{\langle {{#1}} , {{#2}} \rangle}}

\newcommand{\qed}{\(\hfill \square\)}

\newcommand{\Vm}[1]{#1}
\newcommand{\Vv}[1]{\bar{#1}}
\newcommand{\Vd}[1]{\underline{#1}}
\newcommand{\V}[1]{\boldsymbol{#1}}
\newcommand{\M}[1]{\mathtt{{#1}}}
\newcommand{\diff}[1]{{\mathtt{d} {#1}}}

\DeclareMathOperator{\diag}{\mathtt{diag}}
\DeclareMathOperator{\st}{s.t}
\DeclareMathOperator{\p}{\partial}
\DeclareMathOperator{\id}{\mathtt{id}}

\renewcommand{\dim}{\mathtt{dim}}

\newtheorem{lem}{\textbf{Lemma}}

\newtheorem{defn}{\textbf{Definition}}

\title{\LARGE \bf Polynomial mechanics and optimal control}
\author{Akshay Srinivasan$^{1}$, Madhusudhan Venkadesan$^{2}$
  \thanks{*This work was funded by grants to M.~Venkadesan from the Human Frontier Science Program (RGY0091), the National Centre for Biological Sciences and the Simons Foundation. A.~Srinivasan was supported by grants from NSF and DARPA, awarded to Emanuel~Todorov, Department of Computer Science \& Engineering, University of Washington, Seattle.}
  \thanks{$^{1}$ Department of Computer Science \& Engineering, University of Washington, Seattle, WA-98195, USA,
      \tt\footnotesize akshays@cs.washington.edu}
  \thanks{$^{2}$Department of Mechanical Engineering \& Materials Science, Yale University, New Haven, CT 06511, USA,
  	  \tt\footnotesize m.venkadesan@yale.edu}}

\begin{document}
\maketitle
\thispagestyle{empty}
\pagestyle{empty}

\begin{abstract}
  We describe a new algorithm for trajectory optimization of mechanical systems. Our method combines pseudo-spectral methods for function approximation with variational discretization schemes that exactly preserve conserved mechanical quantities such as momentum. We thus obtain a global discretization of the Lagrange-d'Alembert variational principle using pseudo-spectral methods. Our proposed scheme inherits the numerical convergence characteristics of spectral methods, yet preserves momentum-conservation and symplecticity after discretization. We compare this algorithm against two other established methods for two examples of underactuated mechanical systems; minimum-effort swing-up of a two-link and a three-link acrobot.
\end{abstract}

\section{INTRODUCTION}

Trajectory optimization methods are broadly classified into two categories: Direct, and Indirect \cite{betts2010practical}. Direct methods discretize both state and control trajectories to derive a finite-dimensional constrained optimization problem, while indirect methods solve the discretized nonlinear equations resulting from the necessary first-order variational conditions of Euler-Lagrange-Pontryagin \cite{pontryagin1987mathematical}. Direct methods more convenient for the non-specialized practitioner especially because of the difficulty in deriving conditions equivalent to Pontryagin's when the problem involves free parameters or in\-equality constraints \cite{betts2010practical}.

Pseudo-spectral implementations of direct methods have seen increasing use in recent years because of their super-polynomial convergence \cite{fahroo2002direct} \cite{von1992direct} \cite{benson2006direct} \cite{biegler1984solution}. While these methods are applicable for a broad class of systems, they tend however not to preserve certain geometric structures that are associated with conservation laws for mechanical systems. 

The formulation of {\em Discrete-Mechanics and Optimal Control} (DMOC) satisfies the latter requirement by using variational integrators to construct schemes which are both symplectomorphic and momentum-conserving \cite{Ober-blobaum:2011} \cite{junge2005discrete}. These methods however lack the convergence and approximation properties of the former pseudo-spectral methods and only exhibit a fixed-order algebraic convergence.

The results from the theory of Geometric integration \cite{hairer2006geometric} \cite{marsden2001discrete}, are not readily converted to pseudo-spectral discretizations because the polynomial bases only yield quadratures for weighted integrals. Furthermore, the theory only provides statements for the map between initial and end-time values; the global discretization is essentially one step of an ODE integrator.

In this paper, we propose a direct method which incorporates the benefits of both pseudo-spectral and DMOC methods into a single algorithm. We extend the analysis of DMOC, using variations over polynomials, to derive a discrete non-causal analogue of the Euler-Lagrange equation. We also prove that the resulting discretization is both symplectomorphic and momentum-conserving.



In the following sections, we make use of specialized notation which serves the dual purpose of being both intuitive and reasonably precise. The precise definitions can be found in the appendix (\ref{ssec:notation}).

\section{Problem Statement}
%
\subsubsection{Dynamics}
We restrict our attention in this paper to Lagrangian systems. Given the Lagrangian \(\mathcal{L}(\Vm{q}, \Vv{v})\), the corresponding dynamics is generated by the condition that every trajectory-curve of the system \(\Vm{q} : [0, t_f] \rightarrow \mathcal{M}\), satisfy the Lagrange-d'Alembert principle,
\begin{equation}
  \label{eqn:sys}
  \begin{gathered}
    \mathcal{L} : T \mathcal{M} \rightarrow \mathbb{R},\quad  \mathcal{L} |_q : T\mathcal{M}_q \rightarrow \mathbb{R}\; \mbox{is convex.}\\   
    \delta_{ \Vv{\delta q}} \left[\int_0^{t_f} \mathcal{L}(\Vm{q}(t), \Vv{D_t \Vm{q}(t)}) \diff{t}\right] + \int_0^{t_f} \Vd{u}(t) \Vv{\delta q}(t) \diff{t}  = \Vd{p} \Vv{\delta q} |_{0}^{t_f},\\
    \forall \delta \Vv{q} \in (\mathbb{R}[t]_N)^{\dim(\mathcal{M})},\\
    \mbox{where,} \quad \Vd{p}(t) := \p_{v} \mathcal{L} (\Vm{q}(t), \Vv{\dot{q}}(t)).
  \end{gathered}
\end{equation}
Using variational arguments, it can be shown that every such (smooth) solution also satisfies the Euler-Lagrange equations,
\begin{equation}
  D_t \left[ \begin{array}{c}  \Vm{q}(t)\\\p_{v} \mathcal{L} (\Vm{q}(t), \Vv{\dot{q}}(t)) \end{array}\right] = \left[ \begin{array}{c} \Vv{\dot{q}}(t) \\ \p_{q} \mathcal{L} (\Vv{q}(t), \Vv{\dot{q}}(t)) + \Vd{u}(t)\end{array}\right].
\end{equation}

\subsubsection{Control}
The optimal control problem is defined as finding a control sequence which incurs the least cost (application-specific), while respecting the dynamics defined by (\ref{eqn:sys}),
\begin{equation}
  \label{eqn:opt}
  \begin{aligned}
    \Vd{u}^* & = \arg \min_{\Vd{u}} J(\Vm{q_0}, \Vd{u}),\\
    J(\Vm{q_0}, \Vd{u}) &=  \int_{0}^{t_f} l([\Vm{q}(t), \Vv{D_t q(t)}], \Vd{u}(t)) \diff{t} + \\
    & \quad \quad \quad \quad V_f([\Vm{q}(t_f), \Vv{D_t q(t_f)}]) ,\\
    &\st\; \Vm{q}(0) = \Vm{q}_0,\\ &\Vm{q}_{[0, t_f]}, \Vd{u}_{[0, t_f]}\; \mbox{satisfies the conditions of (\ref{eqn:sys}).}
  \end{aligned}
\end{equation}

\section{Preliminaries}

\subsection{Orthogonal Polynomials}
The pseudo-spectral approach uses a finite-degree polynomial basis to describe control and state trajectories. Although our implementation makes use of Chebyshev polynomials, the discretization itself is equally applicable to every orthogonal-polynomial base on a closed interval. We briefly summarize the general theory of orthogonal polynomials.

Let \(\mathbb{R}[t]\) denote the vector space of polynomials with real-coefficients, and let \(\mathbb{R}[t]_n\) denote the \(n\)-dimensional vector space of all polynomials less than or equal to degree \((n - 1)\).

\begin{defn}[Lagrange interpolation polynomial]
  Given {\em grid-points}, \(\{t_i\}_{i = 1}^N\), \(\ell_k\) is defined to be the unique polynomial in \(\mathbb{R}[t]_N\) such that \(\ell_k(t_i) = \delta_{ik}\),
  \begin{equation}
    \ell_k(t; \{t_j\}_{j = 1}^{N}) = \prod_{i \neq k}^N \frac{t - t_i} {t_k - t_i}.
  \end{equation}
\end{defn}

\begin{defn}[Orthogonal polynomials]
Given the inner product on the function space \(L^2([-1, 1])\),
\begin{equation}
  \ipro{f}{g}_w = \int_{-1}^{1} w(t) f(t) g(t) \diff{t}.
\end{equation}
The set of orthogonal polynomials \(\mathcal{P}_w\) corresponding to this metric is an ordered orthogonal basis for \(\mathbb{R}[t]\) under this inner product.
\begin{equation}
  \label{eqn:opoly}
  \begin{gathered}
    \ipro{P_i}{P_j}_w = \gamma_i \delta_{i j},\\
    i > j \iff \deg(P_i) > \deg(P_j), \quad \forall P_i, P_j \in \mathcal{P}_w,
  \end{gathered}
\end{equation}
\end{defn}
where \(\gamma_i \in \mathbb{R}_+\) is the normalization constant for \(P_i\) under the inner product.

Orthogonal polynomials also satisfy---and are numerically computed using---recurrence relations of the form \cite{boyd},
\begin{equation}
  \begin{gathered}
    P_0 = a_0,\quad P_1 = a_1 t\\
    P_n(t) = (a_n t + b_n) P_{n - 1}(t) - c_n P_{n - 2}(t).
  \end{gathered}
\end{equation}

The specific polynomial bases used in our work, Chebyshev and Legendre polynomials, correspond to the weights \(w(t) = \frac{1}{\sqrt{1 - t^2}}\) and \(w(t) = 1\), respectively.

\subsubsection{Gauss quadrature, Pullback inner-product}
The orthogonal polynomial set \(\mathcal{P}_w\) can be used to approximate integrals using the \(N\)-point Gauss quadrature scheme of order \(2N\) \cite{boyd},
\begin{equation}
  \label{eqn:quad}
  \int_{-1}^{1} w(t) f(t) \diff{t} \approx \sum_{k = 1}^{N} w_k f(t_k).
\end{equation}
The quadrature points \(\{t_k\}_{k=1}^N\) are chosen to be the roots of the \(N\)'th orthogonal polynomial. The roots and corresponding quadrature weights can be computed numerically from the eigenvalues and eigenvectors of the Jacobi operator, using the Golub-Welsch algorithm \cite{golubwelsch1969}.

Because the order of the quadrature is \(2N\), it is exact for all \(p \in \mathbb{R}[t]_{2 N}\). As a result, Gauss quadrature defines the discrete inner product \(\ipro{P_i}{P_j}_w^N\) over \(\mathbb{R}[t]_{N}\),
\begin{equation}
  \label{eqn:dpro}
  \begin{aligned}
    \forall i, j < N,&\\
    \ipro{P_i}{P_j}_w &= \int_{-1}^{1} w(t) P_i(t) P_j(t) \diff{t}\\
    &= \sum_{k = 1}^N w_k P_i(t_k) P_j(t_k) = \gamma_i \delta_{i j} \\
    &:= \ipro{P_i}{P_j}_w^N.
  \end{aligned}
\end{equation}
This connection ties collocation using these points, to the Galerkin method \cite{boyd} and, crucially, to the quadrature itself,
\begin{equation}
  \int_{-1}^{1} w(t) p(t) \diff{t} = \ipro{1}{p}_w^N \quad \forall p \in \mathbb{R}[t]_{N}.
\end{equation}


Given polynomials \(p,q \in \mathbb{R}[t]_{N}\), the integral of \(pq\) is given exactly by \(\ipro{p}{q}^N\). Since every \(N\)-point (or basis) representation of a polynomial in \(\mathbb{R}[t]_{N}\) is related to every other by a linear transformation, the pullback of the Legendre inner-product, \(\ipro{\cdot}{\cdot}^{N*}\), preserves \(\ipro{p}{q}\), even though the pullback of the quadrature does not,
\begin{equation}
  \ipro{p}{q} = \ipro{1}{pq}^N = \ipro{p}{q}^{N*} \neq \ipro{1}{pq}^{N*}.
\end{equation}
Seen in the coordinate free sense, the quadrature is generically only of order \(N\). The order \(2N\) quadrature is achieved only when the grid-points are restricted to the roots of \(P_{N}\). The use of the pullback, on the other hand, allows one to represent polynomials over arbitrary grid points, and yet be able to use the discretized \(L^2\) inner-product. The ability to do this is important since collocation on Chebyshev grids gives a \(O(\log(N))\) approximation to the best-uniform-approximation polynomial \cite{sachdeva2013faster}, while that on the Legendre-grid has no such guarantee.

The loss of accuracy in quadrature is not a concern when using Chebyshev polynomials, because the resulting quadrature (associated with Clenshaw-Curtis) is known to be nearly as accurate as Gauss-Lobatto \cite{trefethen2008gauss}. In the general case however it may perform poorly because of Runge's phenomenon \cite{trefethen2000spectral}.

\subsection{Symplectic maps}
The flow generated by Lagrange-d'Alembert systems (\ref{eqn:sys}) are symplectic when the control sequence is fixed in time. Discretizations which preserve this property of the system often exhibit desirable numerical properties \cite{hairer2006geometric}.

There are many equivalent definitions for a diffeomorphism to be symplectic. The following is prevalently used on Darboux coordinates,
\begin{defn}[Symplectic map]
  A diffeomorphism \(f: (p, q) \mapsto (P, Q)\) is said to be symplectic if it leaves the symplectic form, \(J\), invariant,
  \[J = (\p f)^T J \p f, \quad J:= \left[\begin{array}{c c}0 & -\id \\ \id & 0\end{array}\right].\]
\end{defn}

We shall however make use of the following equivalent condition in the upcoming sections \cite[p. 196]{hairer2006geometric},
\begin{lem}[Total differential]
  \label{lem:tdiff}
  A map \(\phi : (\Vd{p}, \Vm{q}) \mapsto (\Vd{P}, \Vm{Q})\) is symplectic if and only if, \(P \Vv{dQ} - \Vd{p} \Vv{dq}\), is a total differential in \(\Vd{dp}, \Vv{dq}\).
\end{lem}

\section{Discrete variational constraint}
\subsection{Discrete Lagrange-d'Alembert}
Let the Lagrangian of the system be given by \(\mathcal{L}(\Vm{q}, \Vv{v})\). Using the notation \(\ipro{\Vd{f}}{\Vv{g}} := \int \Vd{f}(t) \Vv{g}(t) \diff{t}\) for the standard inner product on \(L^2\), (\ref{eqn:sys}) is written as,
\begin{equation}
  \label{eqn:hsys}
  \begin{gathered}
    \delta_{ \Vv{\delta q}} \ipro{\mathcal{L} \circ \Vm{q}}{1} + \ipro{\Vd{u}}{\delta \Vv{q}} = \Vd{p} \Vv{\delta q}|_{0}^{t_f}, \forall \delta \Vv{q},\\
    \mbox{where,}\;\mathcal{L}\circ \Vm{q}(t) := \mathcal{L}(\Vm{q}(t), \Vv{D_t q}(t)),\\
    p(t) := \p_v \mathcal{L}(\Vm{q}(t), \Vv{D_t q}(t)).
  \end{gathered}
\end{equation}

Using the inner-product on Legendre polynomials and collocation points \(\{t_i\}\), this condition is approximated as,
\begin{equation}
  \label{eqn:lsys}
    \delta \ipro{\mathcal{L} \circ \Vm{q}}{1}^{N*} + \ipro{\Vd{u}}{\delta \Vv{q}}^{N*} = \Vd{p} \Vv{\delta q}|_{0}^{t_f},\; \forall \delta \Vv{q} \in (\mathbb{R}[t]_N)^{\dim(\mathcal{M})}.
\end{equation}
Expanding the variation, we find,
\begin{equation}
  \label{eqn:elsys}
    \ipro{\mathcal{L}_q + D^{\dagger} \mathcal{L}_v + \Vd{u}}{\delta \Vv{q}}^{N*} = \Vd{p} \Vv{\delta q}|_{0}^{t_f},\; \forall \delta \Vv{q} \in (\mathbb{R}[t]_N)^{\dim(\mathcal{M})},
\end{equation}
where \(\Vd{p}(t) := \sum_{i = 1}^N \mathcal{L}_{v}^i \; \ell_i(t, \{t_k\})\). 

Next, we prove that the map associated with trajectories satisfying (\ref{eqn:elsys}), is both symplectomorphic and momentum-conserving.

\subsection{Symplecticity, Momentum conservation.}
Consider the dynamics described by (\ref{eqn:sys}). Given the end points \(q_0, q_f \in \mathcal{M}\), there exists a unique polynomial \(q^* \in (\mathbb{R}[t]_N)^{\dim(\mathcal{M})}\) that minimizes the discrete action defined in (\ref{eqn:lsys}), the corresponding momentum polynomial is given by \(\Vd{p^*}(t) := \sum_{i = 1}^N \mathcal{L}_{v}^i \; \ell_i(t, \{t_k\})\). The scheme (\ref{eqn:elsys}) is symplectomorphic if the map \((\Vv{\delta q(\tau')}, \Vd{\delta p(\tau')}) \mapsto (\Vv{\delta q(\tau)}, \Vd{\delta p(\tau)}),\) is symplectic for all \(\tau\).

We define the discretized action over the interval \([\tau', \tau]\), for \(q^*\) to be,
\begin{equation}
  S_{\tau', \tau}(q^*_{\tau'}, q^*_{\tau}) = \ipro{\mathcal{L} \circ \Vm{q^*}}{1}^{N*}_{\tau', \tau}, \quad \tau', \tau \in [0, t_f].
\end{equation}
The pullback of the inner-product in the above equation is generated by linear affine transforms between the intervals \([\tau, \tau']\) and \([0, t_f]\). Note that because the polynomials are defined by their values on the original grid points, the discretized action \(S_{\tau', \tau}\) as defined above, also depends on the values over grid points outside \([\tau', \tau]\). This lack of causality in the action leads to equivalent non-causal notions of symplecticity and momentum conservation.

Taking discrete variations of \(S_{\tau', \tau}\) around \(\Vm{q^*}\),
\begin{equation}
  \label{eqn:svar}
  \delta{S}_{\tau', \tau} = \ipro{\mathcal{L}_q  + \Vd{u}}{ \delta \Vv{q}}_{\tau', \tau}^{N*} + \ipro{\mathcal{L}_v}{ D \delta \Vv{q}}_{\tau', \tau}^{N*}.
\end{equation}





Exploiting the fact that the inner-product, \(\ipro{\mathcal{L}_v}{ D \delta \Vv{q}}_{\tau', \tau}^{N*}\), is exact on \(\mathbb{R}[t]_N\),
\begin{equation}
  \begin{aligned}
    \ipro{\Vd{p}}{D \delta \Vv{q}}^{N*}_{\tau', \tau} &= \int_{\tau'}^{\tau} \Vd{p}(t) D_t \delta \Vv{q}(t) \diff{t},\\
    &= \Vd{p} \delta \Vv{q} |_{\tau'}^{\tau} - \int_{\tau'}^{\tau} D_t \Vd{p}(t) \delta \Vv{q}(t) \diff t,\\
    &= \Vd{p} \delta \Vv{q} |_{\tau'}^{\tau} - \ipro{D \Vd{p}}{ \delta \Vv{q}}^{N*}_{\tau', \tau}.
  \end{aligned}
\end{equation}

Hence,
\begin{equation}
  \label{eqn:svar1}
  \delta{S}_{\tau', \tau} = \ipro{\mathcal{L}_q  + \Vd{u} - D \mathcal{L}_v}{\delta \Vv{q}}^{N*}_{\tau', \tau} + \Vd{p} \delta \Vv{q} |_{\tau'}^{\tau}.
\end{equation}

For the constraint (\ref{eqn:elsys}) to be satisfied, we require,
\begin{equation}
  \label{eqn:pcons}
  \begin{gathered}
  \delta{S}_{0, t_f} := \ipro{\mathcal{L}_q + \Vd{u} - D \mathcal{L}_v}{\delta \Vv{q}}^{N*} + \Vd{p} \delta \Vv{q} |_{0}^{t_f} = \Vd{p} \delta \Vv{q} |_{0}^{t_f},\\
  \forall \delta \Vv{q} \in (\mathbb{R}[t]_N)^{\dim(\mathcal{M})},\\
  \Rightarrow (\mathcal{L}_q + \Vd{u} - D \mathcal{L}_v) = \Vd{0}.
  \end{gathered}
\end{equation}

It follows hence from (\ref{eqn:svar1}) that, \(\delta{S}_{\tau', \tau} = \Vd{p} \delta \Vv{q} |_{\tau'}^{\tau}, \forall \tau', \tau\). This condition along with
Lemma-\ref{lem:tdiff} proves that the discretization (\ref{eqn:elsys}) is symplectic.  \qed

\subsection{Algorithm}
In the collocation ``co-ordinates'', let the metric tensor corresponding to the Legendre inner-product be given by \(G\), the Lagrange derivative matrix on the grid by \(\M{D}\), and the dual forms evaluating the polynomial at \(t = 0, t_f\) by \(L(0), L(t_f)\) respectively. Then the condition (\ref{eqn:elsys}) becomes,
\begin{equation}
  G (\mathcal{L}_q + \Vd{u}) + (\M{D}^T G - [L^T(t_f) L(t_f) - L^T(0) L(0)])  \mathcal{L}_v = 0,
\end{equation}
where we use the fact that \(\M{D}^{\dagger} := G^{-1} \M{D}^T G\).

Incorporating all the constraints from (\ref{eqn:opt}), the optimal control problem can now be approximated as the following finite-dimensional nonlinear optimization problem,
\begin{equation}
  \label{eqn:fopt}
  \begin{gathered}
    \min_{\Vm{q}, \Vd{u}} \V{1}^T G \; \V{l},\\
  G (\mathcal{L}_q + \Vd{u}) + (\M{D}^T G - [L^T(t_f) L(t_f) - L^T(0) L(0)])  \mathcal{L}_v = 0.\\
    L(0) \Vm{q} = \Vm{q}_0,\\
    \mbox{where} \; \V{l}[i] = l(\Vm{q}(t_i), (\M{D} q)(t_i), u(t_i)).
  \end{gathered}
\end{equation}

\section{Numerical examples}
\begin{figure*}[t!]
  \centering
  \includegraphics[height=9cm]{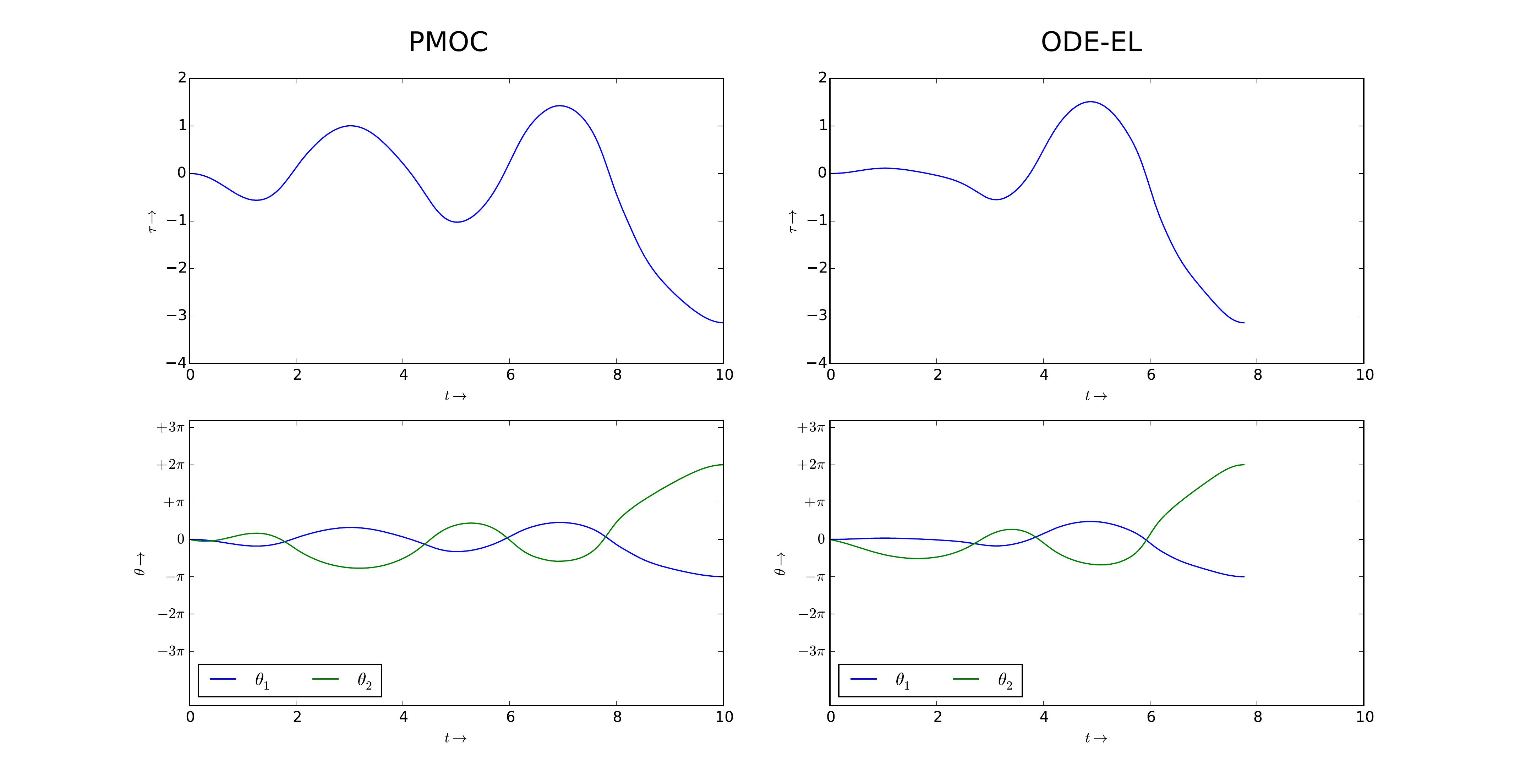}
  \caption{Acrobot: Locally optimal solutions in \(\mathbb{R}[t]_{64}\) found by (Left) PMOC \& (Right) ODE-EL; (Top) Optimal control sequence; (Bottom) Corresponding trajectory found by the optimizer. The solution found by ODE-EL takes a shorter time to swing-up and also incurs a higher cost.}
  \label{fig:pmoca}
\end{figure*}

We illustrate the performance of the algorithm described in this paper, using the classical Acrobot, and a 3-link analogue that we call the {\em 3crobot}: a 3-link pendulum with free pivots on all but the last joint that has a torque actuator.  The models in both-cases are non-dimensionalized using the mass \& length of the first link, and time expressed in units such that \(\mathtt{g} = 1.0\).

The common goal in both control problems is to start off from the lowest-energy state and swing up to the upright position, while minimizing \(\int_{0}^{t_f} |\tau|^2 \diff{t}\), and optimizing \(t_f \in [0, 10]\). The total duration for the optimization is bounded, but not exactly specified, thereby introducing a parameter to be optimized in addition to the control sequence.

The scheme presented in this paper is abbreviated by PMOC. The acronym DAE-EL refers to the pseudo-spectral discretization of the Euler-Lagrange condition : \(D \mathcal{L}_v = \mathcal{L}_q + \Vd{u}\). ODE-EL refers to the pseudo-spectral discretization of the resultant first-order ODE from Euler-Lagrange: \(D \Vv{\dot{q}} = (\mathcal{L}_{vv})^{-1} (\mathcal{L}_q + \Vd{u} - \mathcal{L}_{qv} \Vv{\dot{q}})\). The problems are discretized using Chebyshev polynomials, and the resulting nonlinear programs (\ref{eqn:fopt}) are solved using SNOPT \cite{gill2005snopt} \footnote{Our implementation assumes all the constraints to be nonlinear; an assumption that impacts all the schemes considered here equally.}.

\subsection{Acrobot}
Taking inspiration from \cite{tedrake2009underactuated}, we obtain the initial guess for the problem by using a sinusoidal waveform for \(\tau(t)\). The number of major iterations taken by the SQP solver are listed in Table \ref{tbl:maja}.

We see that both PMOC and ODE-EL, converge to locally optimal solutions, but DAE-EL fails to find a feasible solution. PMOC converges with fewer major iterations than ODE-EL.

\subsection{3crobot}
\begin{figure}[!t]
  \centering
  \includegraphics[height=5cm]{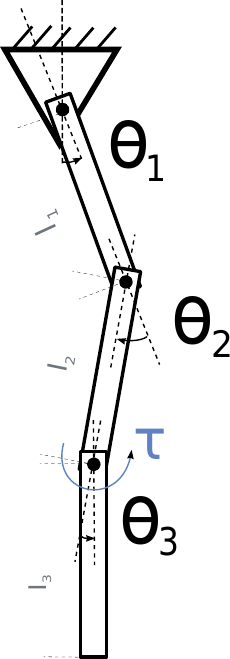}
  \caption{3crobot: The system is composed of 3 links of lengths \(1.0, l_2, l_3\) respectively, connected together by pivot joints as shown in the figure. An actuator situated on the joint farthest from the ground, can apply arbitrary torques \(\tau\) on the joint. The control task is to start at rest from \(\theta_{\{1, 2, 3\}} = 0\) and swing-up to \(\theta_1 = \pi, \theta_{\{2, 3\}} = 0\) (or equivalent co-ordintates) while minimizing the cost \(\int_{0}^{t_f} | \tau |^2 \diff{t},\; t_f \in [1, 10] \mathtt{s}\).}
  \label{fig:3crobot}
\end{figure}

The schematic for the 3crobot is illustrated in Fig.\ref{fig:3crobot}. We consider two versions of the control problem: one where the lengths of the links, \(l_2, l_3\) are fixed at \(0.5\) each, and the other where it is required to find the optimal values for \(l_2, l_3\), such that \(l_2 + l_3 = 1.0\). Such situations arise in coupled optimal control and design problems.

The initial (infeasible) guess was found by applying a constant torque around \(\theta_1\), while the remaining joints were stabilized around \(0\) with a proportional controller. The number of major iterations taken by the SQP solver are listed in Table \ref{tbl:maj3}.

In both cases, PMOC finds a local optimum faster than ODE-EL. ODE-EL had trouble satisfying the feasibility tolerance and exceeded the resource limit, but it appears to have landed in a neighborhood of the solutions found by PMOC. DAE-EL failed to converge in both cases.

\begin{figure*}[t!]
  \centering
  \includegraphics[height=9cm]{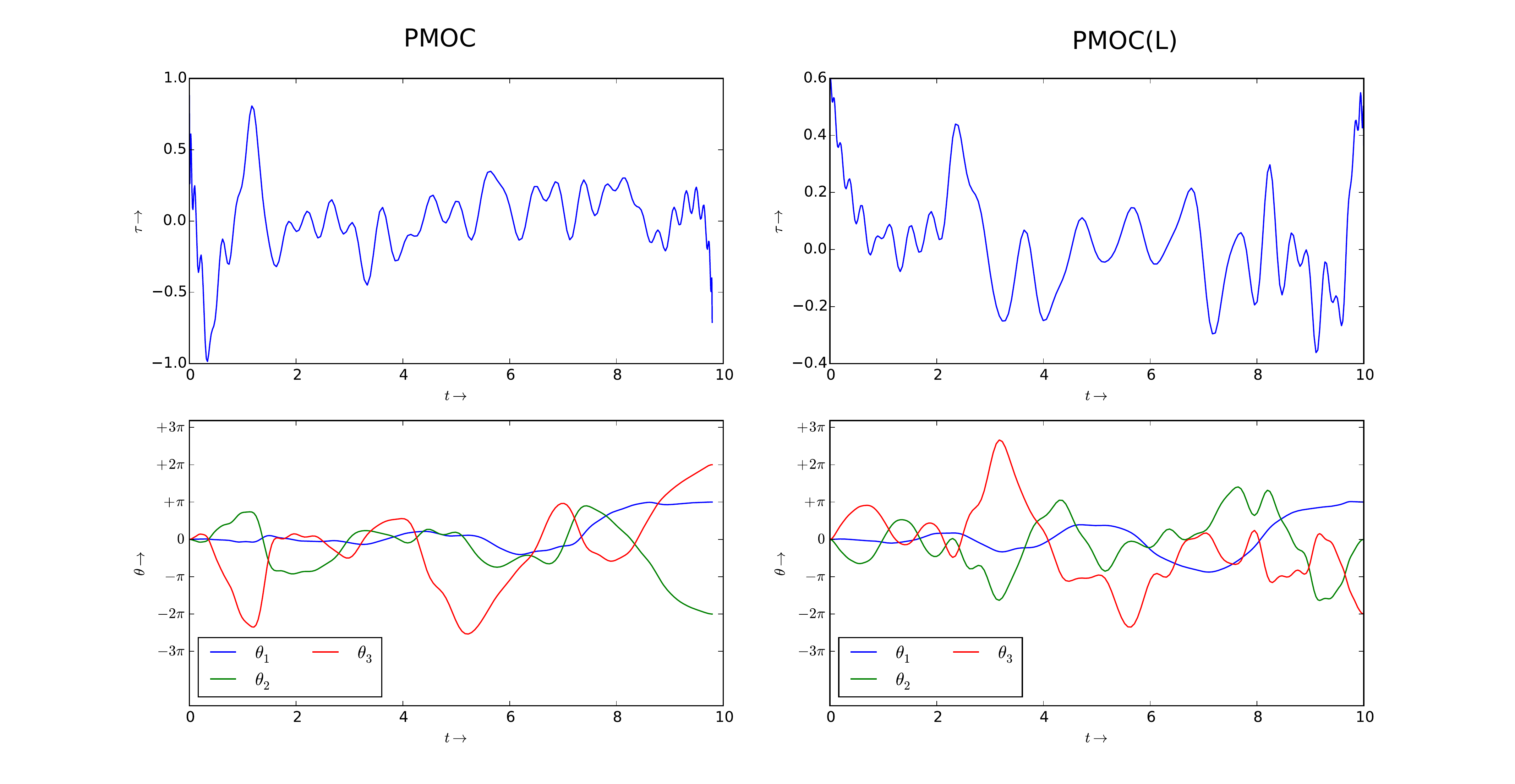}
  \caption{3crobot: Locally optimal solutions in \(\mathbb{R}[t]_{64}\) found by PMOC; (Left) Solution for fixed \(l_2 = l_3 = 0.5\); (Right) Solution for link-lengths s.t \(l_2 + l_3 = 1\): optimal parameter was found to be \((l_2^*, l_3^*) \approx (0.3,0.7)\); (Top) Optimal control sequence; (Bottom) Corresponding trajectory found by the optimizer.}
  \label{fig:pmoc3}
\end{figure*}

\begin{table}[!t]
  \centering
  \begin{tabular}{| c | c | c |}
    \hline
    Algorithm & Major iterations & Cost\\
    \hline
    PMOC & 218 & 0.63\\
    DAE-EL & No feasible solution found & -\\
    ODE-EL & 688 & 0.80 \\
    \hline
  \end{tabular}
  \caption{Performance on the Acrobot problem}
  \label{tbl:maja}
\end{table}

\begin{table}[!t]
  \centering
  \begin{tabular}{| c | c | c |}
    \hline
    Algorithm & Major iterations & Cost\\
    \hline
    PMOC & 498 & 0.61\\
    DAE-EL & Singular basis & -\\
    ODE-EL & \(> 1758\) & - \\
    \hline
    PMOC (\(l\)) & 358 & 0.31\\
    DAE-EL (\(l\)) & Singular basis & -\\
    ODE-EL (\(l\)) & \(> 1758\) & - \\
    \hline
  \end{tabular}
  \caption{Performance on the 3crobot problem}
  \label{tbl:maj3}
\end{table}






\section{Summary}

Optimal control problems are generally non-convex, and underactuated problems of the kind presented in this paper further accentuate the difficulty of finding the global or even a local minimum. Every smooth optimal control algorithm is susceptible to local optima, and this is partially addressed in practice using multiple starts. The challenge however is in designing algorithms that, more often than not, find feasible solutions. As seen from our numerical examples, even established methods sometimes fail to find feasible solutions.

We show that the algorithm that we propose enjoys computational advantages over the other candidates considered here; faster convergence and consistently finds feasible solutions. This comparison is by no means exhaustive, and only meant to be representative of the current state of the art in trajectory optimization. We also prove that the discretization used in our algorithm is both symplectic and momentum-conserving, and incorporates the beneficial aspects of pseudo-spectral methods. While our algorithm appears to also find ``better'' optima, we caution the reader against paying heed to this aspect of the results. With judicious use of multiple starts, it is possible that other algorithms may find comparable optima.

Performance of any of these methods is likely to be problem-dependent. For example, because there exist many different costs that generate the same optimal behavior of the system \cite{ng2000algorithms}, the cost function itself could presumably be tuned so as to favor the performance of one algorithm over the others. With the exception of such specially tuned cost functions, we find that PMOC is especially effective on complex mechanical systems with commonly used cost functions such as torque-squared, minimum-time and so on. 

The use of variational integrators has been extensively pursued in the DMOC literature \cite{Ober-blobaum:2011} \cite{junge2005discrete} \cite{marsden2001discrete}. Unlike DMOC, the algorithm presented in this paper sacrifices causality in order to better approximate the dynamics using pseudo-spectral methods. While this leads to super-polynomial (as opposed to fixed-order algebraic) convergence rate requiring fewer grid points, it also leads to dense Jacobians and increased sensitivity to discontinuities. This trade-off is reminiscent of finite-element versus spectral methods, and suggests the future development of adaptive-variational schemes resembling hp-adaptive pseudo-spectral methods for optimal control \cite{darby2011hp}.

The poor performance of DAE-EL in our examples underscores the lack of our present understanding on how geometric discretizations affect numerical methods for optimal control. It is known that ODE-stepping schemes based on the Gauss-Lobatto quadrature are both symplectic \& momentum-conserving when working in the DAE-EL (Hamiltonian) form \cite[p. 192]{hairer2006geometric}. Similar grid densities between Chebyshev and Legendre basis predisposed us to expect similar performance for DAE-EL and PMOC, but this clearly does not seem to be case for these examples (see Table \ref{tbl:maja}). This behaviour can partly be attributed to the ill-conditioning of the derivative matrix \(\M{D}_{ij}\), a property which is ameliorated in PMOC by the use of the conjugate operator \(D^{\dagger}\). The contrast in performance also seems to make the case for symplectic discretizations, since DAE-EL like PMOC, is also momentum-conserving.

The results from this paper suggest that for a discretization of a given order, those that are geometry preserving offer advantages in terms of the rates and region of convergence for optimal control problems where finding feasible solutions is challenging. Future investigations will examine how the choice of the polynomial basis for pseudo-spectral interpolation affects the performance of the numerical method.

\bibliographystyle{IEEEtran}
\bibliography{IEEEfull,references}

\begin{thebibliography}{10}
\providecommand{\url}[1]{#1}
\csname url@rmstyle\endcsname
\providecommand{\newblock}{\relax}
\providecommand{\bibinfo}[2]{#2}
\providecommand\BIBentrySTDinterwordspacing{\spaceskip=0pt\relax}
\providecommand\BIBentryALTinterwordstretchfactor{4}
\providecommand\BIBentryALTinterwordspacing{\spaceskip=\fontdimen2\font plus
\BIBentryALTinterwordstretchfactor\fontdimen3\font minus
  \fontdimen4\font\relax}
\providecommand\BIBforeignlanguage[2]{{%
\expandafter\ifx\csname l@#1\endcsname\relax
\typeout{** WARNING: IEEEtran.bst: No hyphenation pattern has been}%
\typeout{** loaded for the language `#1'. Using the pattern for}%
\typeout{** the default language instead.}%
\else
\language=\csname l@#1\endcsname
\fi
#2}}

\bibitem{betts2010practical}
J.~T. Betts, \emph{Practical methods for optimal control and estimation using
  nonlinear programming}.\hskip 1em plus 0.5em minus 0.4em\relax Siam, 2010,
  vol.~19.

\bibitem{pontryagin1987mathematical}
L.~S. Pontryagin, \emph{Mathematical theory of optimal processes}.\hskip 1em
  plus 0.5em minus 0.4em\relax CRC Press, 1987.

\bibitem{fahroo2002direct}
F.~Fahroo and I.~M. Ross, ``Direct trajectory optimization by a chebyshev
  pseudospectral method,'' \emph{Journal of Guidance, Control, and Dynamics},
  vol.~25, no.~1, pp. 160--166, 2002.

\bibitem{von1992direct}
O.~von Stryk and R.~Bulirsch, ``Direct and indirect methods for trajectory
  optimization,'' \emph{Annals of Operations Research}, vol.~37, no.~1, pp.
  357--373, 1992.

\bibitem{benson2006direct}
D.~A. Benson, G.~T. Huntington, T.~P. Thorvaldsen, and A.~V. Rao, ``Direct
  trajectory optimization and costate estimation via an orthogonal collocation
  method,'' \emph{Journal of Guidance, Control, and Dynamics}, vol.~29, no.~6,
  pp. 1435--1440, 2006.

\bibitem{biegler1984solution}
L.~T. Biegler, ``Solution of dynamic optimization problems by successive
  quadratic programming and orthogonal collocation,'' \emph{Computers \&
  chemical engineering}, vol.~8, no.~3, pp. 243--247, 1984.

\bibitem{Ober-blobaum:2011}
S.~Ober-bl{\"o}baum, O.~Junge, and J.~E. Marsden, ``Discrete mechanics and
  optimal control: An analysis,'' \emph{ESAIM: Control, Optimisation and
  Calculus of Variations}, vol.~17, pp. 322--352, 2011.

\bibitem{junge2005discrete}
O.~Junge, J.~E. Marsden, and S.~Ober-Bl{\"o}baum, ``Discrete mechanics and
  optimal control,'' in \emph{IFAC Congress, Praha}, 2005.

\bibitem{hairer2006geometric}
E.~Hairer, C.~Lubich, and G.~Wanner, \emph{Geometric numerical integration:
  structure-preserving algorithms for ordinary differential equations}.\hskip
  1em plus 0.5em minus 0.4em\relax Springer, 2006, vol.~31.

\bibitem{marsden2001discrete}
J.~E. Marsden and M.~West, ``Discrete mechanics and variational integrators,''
  \emph{Acta Numerica}, vol.~10, no.~1, pp. 357--514, 2001.

\bibitem{boyd}
J.~P. Boyd, \emph{Chebyshev and Fourier Spectral Methods}.\hskip 1em plus 0.5em
  minus 0.4em\relax Dover, 2001.

\bibitem{golubwelsch1969}
G.~H. Golub and J.~H. Welsch, ``Calculation of gauss quadrature rules,''
  \emph{Mathematics and Computation}, vol.~23, pp. 221--230, 1969.

\bibitem{sachdeva2013faster}
S.~Sachdeva and N.~K. Vishnoi, ``Faster algorithms via approximation theory,''
  \emph{Theoretical Computer Science}, vol.~9, no.~2, pp. 125--210, 2013.

\bibitem{trefethen2008gauss}
L.~N. Trefethen, ``Is gauss quadrature better than clenshaw-curtis?''
  \emph{SIAM review}, vol.~50, no.~1, pp. 67--87, 2008.

\bibitem{trefethen2000spectral}
------, \emph{Spectral methods in MATLAB}.\hskip 1em plus 0.5em minus
  0.4em\relax Siam, 2000, vol.~10.

\bibitem{gill2005snopt}
P.~E. Gill, W.~Murray, and M.~A. Saunders, ``Snopt: An sqp algorithm for
  large-scale constrained optimization,'' \emph{SIAM review}, vol.~47, no.~1,
  pp. 99--131, 2005.

\bibitem{tedrake2009underactuated}
R.~Tedrake, ``Underactuated robotics: Learning, planning, and control for
  efficient and agile machines course notes for mit 6.832.''

\bibitem{ng2000algorithms}
A.~Y. Ng, S.~J. Russell, \emph{et~al.}, ``Algorithms for inverse reinforcement
  learning.'' in \emph{Icml}, 2000, pp. 663--670.

\bibitem{darby2011hp}
C.~L. Darby, W.~W. Hager, and A.~V. Rao, ``An hp-adaptive pseudospectral method
  for solving optimal control problems,'' \emph{Optimal Control Applications
  and Methods}, vol.~32, no.~4, pp. 476--502, 2011.

\end{thebibliography}

\section*{Appendix}

\subsection{Notation} \label{ssec:notation}
We use \(\ipro{\cdot}{\cdot}\) to denote the standard inner-product on \(L^2\). An N-point discretization via Gauss-Legendre quadrature is represented by \(\ipro{\cdot}{\cdot}^{N}\). A \('*'\) symbol in the super-script : \(\ipro{\cdot}{\cdot}^{N *}\), is used to emphasize the grid-points at which the inner-product is computed; this being important for reasons of approximation. We denote by \(D^{\dagger}\) the conjugate linear operator under the inner product: \(\ipro{f}{Dg} = \ipro{D^{\dagger} f}{g}\). The matrix corresponding to a inner-product in some particular basis is termed the ``Metric tensor'', and denoted by \(G\).

We employ a specialized notation for denoting vectors and duals over the state space of the system. A dual form is marked by a bar underneath the symbol: '\(\Vd{y}\)', whilst a vector with one above it: '\(\Vv{x}\)'. The canonical pairing between a vector and a dual is denoted without special operators: \(\Vd{y} \Vv{x} = \Vv{x} \Vd{y}\). We also assume that operators on the function space act element-wise on a ``stack'' of elements - derivatives of a vector of polynomials, for instance. This also implies that the canonical pairing commutes with inner-products: \(\ipro{\Vv{f}}{\Vd{g}} = \sum_{i} \ipro{\Vv{f}_i}{\Vd{g}_i}\). Partial derivatives of functions are assumed to be dual vectors (or ``stacks'' thereof) throughout the paper.

Generic vectors without special connotation or type will be represented in  bold: '\(\V{x}\)'. Matrices (and metric tensors) are denoted in capitals '\(\M{B}\)'. Generic vectors and matrices will appear in co-ordinate bound expressions, whilst the dual/tangent vectors. Note that unlike the mechanics literature, ``co-ordinates'' here refers to the polynomial base (or equivalently grid-points), and not the state space of the system.


\subsection{Computing the Pullback}
Let the collocation points, weights, and norm-squares (defined in (\ref{eqn:opoly}), (\ref{eqn:dpro})) corresponding to the N-point Gauss-Lobatto quadrature be \(\{t_i^l\}, \{w_i^l\}, \{\gamma_i^l\}\) respectively, and let those corresponding to the Orthogonal polynomial of interest \(\mathcal{B}\), be \(\{t_i^b\}, \{w_i^b\}, \{\gamma_i^b\}\). Given polynomials \(x,y \in \mathbb{R}[t]_{N}\) let,
\begin{equation}
  \begin{aligned}
    \V{x}^b_i &:= x(t_i^b), \V{x}^l_i := x(t_i^l), \V{y}^b_i := y(t_i^b), \V{y}^l_i := y(t_i^l),\\
    &\quad \quad \quad \M{B}_{ij} := B_j(t_i^b), \M{B}^l_{ij} := B_j(t_i^l).\\
    (x, y) &:= \int_{-1}^{1} x(t) y(t) \diff{t} = \sum_{i} x(t_i^l) y(t_i^l) w_i^l
    \\ & = (\V{x}^l)^T \diag(\V{w}^l) \V{y}^l.
  \end{aligned}
\end{equation}
Using the orthogonality of the basis polynomials (\ref{eqn:quad}), we note that,
\begin{equation}
  \begin{gathered}
    \V{p}^b_i := p(t_i^b), \V{p}^l_i := p(t_i^l), \\
    \V{p}^l = \M{B}^l \diag(1./\V{\gamma}^b) \M{B}^T \diag(\V{w}^b) \V{p}^b,
  \end{gathered}
\end{equation}
Hence,
\begin{equation}
  \begin{gathered}
    (p, q) = \V{p}^l \diag(\V{w}^l) \V{q}^l = (\V{p}^b)^T \M{A}^T \diag(\V{w}^l) \M{A} \V{q}^b,\\
    \M{A} := \M{B}^l \diag(1./\V{\gamma}^b) \M{B}^T \diag(\V{w}^b).
  \end{gathered}
\end{equation}



The pullback of the inner-product is therefore,
\begin{equation}
    G^* = \M{A}^T \diag(\V{w}^l) \M{A}, \quad \M{A} := \M{B}^l \diag(1/\V{\gamma^b}) \M{B}^T \diag(\V{w^b}).
\end{equation}


\end{document}